\documentclass[sigplan,10pt]{acmart}
\usepackage{booktabs} 
 \usepackage{enumitem}
\usepackage{balance}
\usepackage[linesnumbered,ruled,vlined]{algorithm2e}
\usepackage{listings}
\usepackage{cleveref}
\usepackage{graphicx}

\setcopyright{none} 
\renewcommand\footnotetextcopyrightpermission[1]{} 
\SetKwInput{KwInput}{Input}                
\SetKwInput{KwOutput}{Output}              

\SetCommentSty{commfont}

\begin{document}
\title{PolyScientist: Automatic Loop Transformations Combined with Microkernels for Optimization of  Deep Learning Primitives}

\author{Sanket Tavarageri}
\affiliation{
  \institution{Intel Labs}
}
\email{sanket.tavarageri@intel.com}

\author{Alexander Heinecke}
\affiliation{
  \institution{Intel Labs}
    }
\email{alexander.heinecke@intel.com}

\author{Sasikanth Avancha}
\affiliation{
  \institution{Intel Labs}
    }
\email{sasikanth.avancha@intel.com}

\author{Gagandeep Goyal}
\affiliation{
  \institution{IIT Hyderabad}
    }
\email{cs19mtech01003@iith.ac.in}

\author{Ramakrishna Upadrasta}
\affiliation{
  \institution{IIT Hyderabad}
    }
\email{ramakrishna@iith.ac.in}

\author{Bharat Kaul}
\affiliation{
  \institution{Intel Labs}
    }
\email{bharat.kaul@intel.com}


\begin{abstract}
At the heart of deep learning training and inferencing are computationally intensive primitives such as \emph{convolutions} which 
form the building blocks of deep neural
networks. Researchers have taken two distinct approaches to creating high performance implementations of deep learning kernels, namely, 1) library development exemplified by Intel MKL-DNN for CPUs, 2) automatic compilation represented by the TensorFlow XLA compiler. The two approaches have their drawbacks: even though a custom built library can deliver very good performance, the cost and time of development of the library can be high. Additionally, hand coding of a plethora of operators for performance is not scalable over the long term as more and more deep learning operators get invented. 
Automatic compilation of kernels is attractive but in practice, till date, automatically generated implementations lag expert coded kernels in performance by orders of magnitude.

In this paper, we develop a hybrid solution to the development of deep learning kernels that achieves the best of both worlds: the expert coded \emph{microkernels} are utilized for the innermost loops of kernels that exploit the vector register files, and vector units of modern CPUs effectively, and we use the advanced polyhedral compilation technology to automatically tune the outer loops for performance. 
We design a novel polyhedral model based data reuse algorithm to optimize the outer loops of the kernel. The overall effect of this combined approach will be that 1) the library development effort is reduced to writing of only a small number of tiny kernels that occur commonly in deep learning workloads, and thus library development is made scalable; 2) automatic compilation with the use of expert-coded microkernels will achieve state-of-the art high performance.
Through experimental evaluation on an important class of deep learning primitives namely convolutions, we demonstrate that the approach we develop
attains the same levels of performance as Intel MKL-DNN, a hand coded deep learning library.

\end{abstract}

%
%

%
%

\maketitle

\section{Introduction}
\label{sec:intro}

Deep learning has revolutionized many spheres of human activity, examples of which include, speech recognition \cite{hinton2012deep}, image recognition \cite{krizhevsky2012imagenet,he2016deep}, web search \cite{googleai}, language translation \cite{wu2016google}, conversational artificial intelligence \cite{devlin2018bert} etc. 
Training and inferencing using deep neural networks (DNNs) that lie at the heart of deep learning are computationally intensive tasks.
Because of the importance of deep learning and the need to speed up training and inferencing,
custom chips such as Google TPUs, Intel Nervana chips, NVIDIA TensorCores have been built. On the software side, frameworks such as TensorFlow, PyTorch, Intel MKL-DNN have been 
created to allow data scientists to write high performing deep learning code in 
an efficient manner. 
However, all these frameworks use manually optimized primitives to deliver
high performance.

The proliferation of computer architectures and the invention of a variety of deep learning kernels has brought to the fore the need to generate high performing kernel implementations that undergird deep learning automatically.
Existing automatic compilation, and auto-tuning techniques  \cite{uday08pldi,Kong:2019:MTM:3314221.3314653,tavarageri2013adaptive,renganarayanan2007parameterized,
darte2014parametric,baskaran2010parameterized,hartono2009parametric,
tavarageri2010parametric,chen2018learning,baghdadi2019tiramisu,chung2004using,
chen2007model,tiwari2009scalable} are inadequate to match
the performance of hand-tuned library implementations -- later on in the paper we show
that the state-of-the-art compiler generated code can lag library implementations on CPUs
by as much as \textasciitilde 10X or more.
The main reason for the failure of automatic compiler techniques in achieving very high performance levels needed is that
the sophistication in the CPU microarchitecture has increased over successive generations of CPUs (data prefetching, speculative execution, vector units, deep memory hierarchies, complex cache replacement algorithms etc). Consequently, 
the cost functions used to optimize code are unable to capture the nitty-gritties of the underlying architectures, and therefore are unable to derive the most effective execution schedules.
Auto-tuning is an alternative approach where one explores a large number of
program variants and selects the best performing version, sidestepping the 
complexity of deriving a cost function that adequately models the 
intricacies of the underlying hardware architecture. However, auto-tuning is
expensive and furthermore, it may fall short of manually created library in performance \cite{baghdadi2019tiramisu}.

\begin{figure*}[h!]
\centering
\begin{minipage}{0.49\textwidth}
\centering
\includegraphics[scale=0.5]{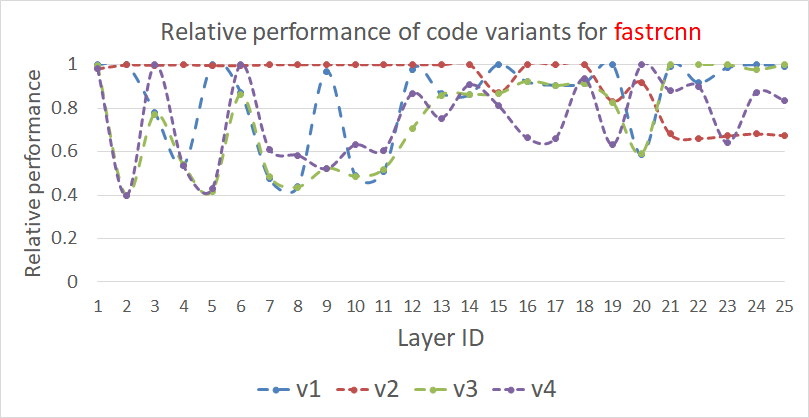}
\caption{Performance of four code variants on layers of \emph{fastrcnn} CNN model on a 28-core Intel Cascade Lake server}
\label{fig:fastrcnn_motivation}
\end{minipage}
\begin{minipage}{0.49\textwidth}
\centering
\includegraphics[scale=0.5]{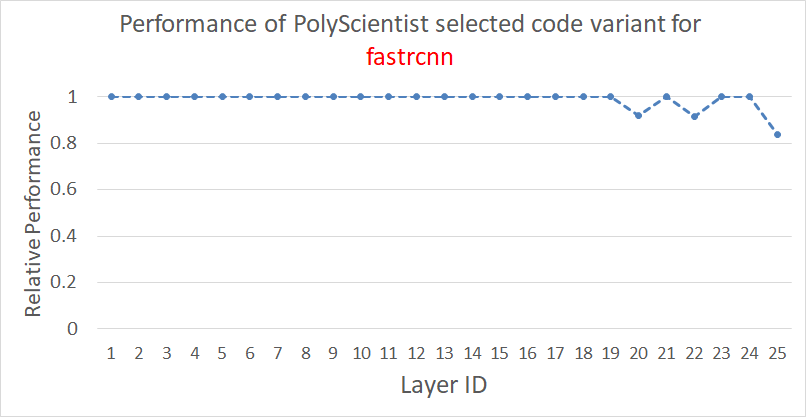}
\caption{Performance of four code variants on layers of \emph{fastrcnn} CNN model on a 28-core Intel Cascade Lake server}
\label{fig:fastrcnn_motivation_polyscientist}
\end{minipage}
\end{figure*}

In this paper, we develop an approach that combines manually optimized
\emph{microkernels} with automatic compilation techniques.
The inner most loops of kernels will be replaced with \emph{microkernels}
and the outer loops will be automatically optimized.
It has been shown that 95\% of all deep learning applications running in the
data centers today have a recurring pattern in the inner most loops, namely blocked
matrix multiplication \cite{jouppi2017datacenter,georganas2019high}.
Therefore, if a microkernel such as that of matrix multiplication that exploits the vector units of modern microprocessors
effectively is used for the inner most loops, and the outer loops are compiled
to use the multi-level caches of the hardware system well, then the resulting
performance will be very high and will be competitive with that of a hand-tuned library.

We use a code-generator to create a number of program variants - $n$ in number -  for a given program. 
The generated code variants are then analyzed by our novel data reuse algorithm -- 
\emph{PolyScientist} to characterize their cache behavior.
We have developed a \emph{relative ranking} algorithm which ranks the $n$ variants based on their potential performance. The top $k$ variants are selected and are run on the target hardware
and the best performing program version is discovered.
Thus, \emph{PolyScientist} narrows down the number of variants to actually run on the target architecture
from $n$ to a small, manageable $k$ ($k << n$). 
Through our experimental evaluation on convolutions of a range of popular and the state-of-the-art image recognition models, we show that
the top $k$ variants picked are some of the best performing variants, and
the realized performance is close to and in many cases, higher than that of Intel MKL-DNN library \cite{intelmkldnn}, a hand-tuned library for deep learning kernels.

The contributions of the paper are the following:
\begin{itemize}
 \item To the best of our knowledge, this is the first work that examines
  the use of microkernels and automatic compilation techniques in an integrated
  manner.
  \item A novel cache data reuse analysis to characterize a loop nest's behavior with respect to a multi-level cache hierarchy is presented.
   \item We describe a \emph{relative ranking} heuristic that takes the compiler generated
   statistics and the system parameters, i.e., cache sizes and ranks the program
   variants based on their potential performance.
   \item We conduct extensive experiments to demonstrate the efficacy of the developed
   system in being able to produce high performance on deep learning kernels.
\end{itemize}

The rest of the paper is organized as follows.
The overall architecture of the \emph{PolyScientist} system is presented in Section \ref{sec:alltogether}.
Section \ref{sec:background} describes preliminary concepts that will 
be used later on. 
In Section \ref{sec:reuse}, we develop the cache data reuse analysis and a poly-ranking
system to rank the candidate program variants in terms of performance.
Section \ref{sec:experiments} details the experimental evaluation conducted.
The related work is discussed in Section \ref{sec:related}
while Section \ref{sec:conclusion} presents the conclusion of this work.

\section{Motivation}
\label{sec:motivation}
The deep learning primitives are computationally intensive and most of the 
neural network training and inferencing time is spent in them.
However, for different layers of the deep neural networks, the optimizations
 (e.g., loop order, tile sizes in tiled code etc) that need to applied are different.
 Using a version of the code optimized for one layer of a neural network for all
 others would yield poor performance.
 It is this need for custom design for different layers of neural networks (the number of layers in a deep neural network can be large) is what makes generating efficient code for
 deep learning primitives a challenging problem.
 To illustrate the need for such tailored loop optimizations we consider 
 the convolution layers of the Fast R-CNN model  \cite{girshick2015fast},
 one of the leading image recognition CNN models.
 We generate four variants of convolution code which differ only in the loop order and the rest of the loop structure remains the same for all of them (more details are
 provided in \S \ref{sec:experiments}) and measure performance on 
 a 28-core Intel(R) Xeon(R) Platinum 8280 (a.k.a Cascade Lake) CPU server.
 Figure \ref{fig:fastrcnn_motivation} shows the normalized performance
 of the code variants on 25 convolution layers of Fast R-CNN: 
 the performance is normalized with respect to the highest performing code
 among the four variants.

\begin{figure*}[t!]
\centering
\includegraphics[scale=0.65]{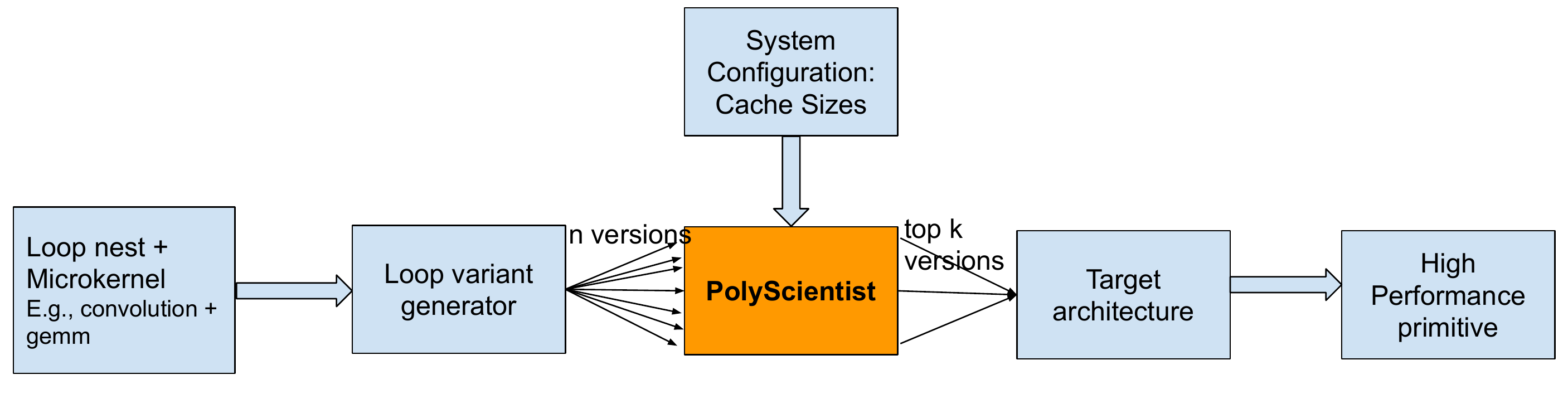}
\caption{The PolyScientist system}
\label{fig:system}
\end{figure*}

From Figure \ref{fig:fastrcnn_motivation}, we observe that the performance of 
different versions of the code varies widely from layer to layer. 
A convolution layer differs from another convolution layer in problem sizes -- 
image sizes, channel widths,
filter sizes, strides, and padding.
The code version \emph{v2} has high performance from layer 1 through 19,
and is subpar from layer 20 to 27. 
The efficiencies of the other versions viz., \emph{v1}, \emph{v3}, and \emph{v4}
are much more widely varying.
Using the methodology detailed in the rest of the paper, the compiler technology
we have developed -- the \textsf{PolyScientist}, we are able to effectively analyze
the four code variants and pick the best performing variant whose performance
is shown in Figure \ref{fig:fastrcnn_motivation_polyscientist}.
The performance achieved by PolyScientist picked code is close to the highest
performance among the four variants for all 25 layers of Fast R-CNN.
Thus using PolyScientist, using compile-time static analysis alone, we will be able to automatically identify and apply
the loop optimizations required for each layer of a deep neural network in order to 
achieve the best performance.

\section{Overall System Architecture}
\label{sec:alltogether}

Figure \ref{fig:system} shows the overall system design. The input to the system is,
the loop nest to be optimized -- $\mathcal{L}$ along with the microkernel
 that forms the inner-most loops.
The loop based specification of the microkernel -- $\mathcal{M}$ is substituted in the code for further analysis. The resulting loop structure -- $\mathcal{L'}$ is regular.
The code generator takes the loop nest $\mathcal{L'}$ and generators a large number
of program variants while keeping the inner most loops that correspond to 
$\mathcal{M}$ intact.
For each generated code variant, the working set sizes are computed as described in 
\S\ref{sec:wscompute}.
The statistics calculated for all the variants are then input to the poly-ranking algorithm
described in \S\ref{sec:wholepolyranking} and it picks the top $k$ best performing versions.
The original microkernels are inserted back into the code of the $k$ picks.
The top performing variants selected analytically are now run on the target
architecture and 
and best performing code among the $k$ loop nests is determined.

\textbf{Microkernel Specification}
The microkernel function call is annotated with a pragma compiler directive which contains the loop-based functionally equivalent code.
The microkernel function call is substituted with the loop based code for the compiler
analysis in a pre-processing pass. When the cache data reuse analysis and ranking of
the code variants are done, in a post-processing pass, the loop-based inner most loops
are replaced with the call to the microkernel.

\section{Preliminaries}
\label{sec:background}

\subsection{Notation}
We use the polyhedral model \cite{feautrier1996automatic}, which is an advanced mathematical framework to reason about dependences and loop transformations, to develop our data reuse algorithm. 
We use the Integer Set Library \cite{verdoolaege2010isl} for performing polyhedral operations in this work and we use the same notation as used in ISL to elucidate the concepts and the algorithm.
The matrix multiplication code shown in Figure \ref{matmulcode} will be used to illustrate the workings of the data reuse analysis.

\begin{figure}
\begin{lstlisting}[language=C,basicstyle=\normalsize,frame=bottomline]
for (i = 0; i < M; i++) {
    for (j = 0; j < N; j++) {
	    for (k = 0; k < K; k++) {
    	    C[i][j] += A[i][k] * B[k][j];
        }
    }
}
\end{lstlisting}
\caption{Matrix multiplication code}
\label{matmulcode}
\end{figure}

\paragraph{Sets}
A set is a tuple of variables $x_i$s along with a collection of constraints $c_k$s defined on the tuple variables. 
$$ s = \{ [x_1, \dots , x_n] : c_1 \land \dots c_m \} $$

The iteration spaces of loop nests are represented by sets. The iteration space of the loop in Figure \ref{matmulcode} is defined as the following set.
$$ I = \{ S[i, j, k] : 0 <= i < M \land 0 <= j < N \land 0 <= k < K \} $$

\paragraph{Relations}
A relation is a mapping from input tuple variables $x_i$s to output tuple variables $y_j$s.
In addition, a set of constraints $c_k$s can be defined for a relation place constraints on the input/output tuple variables. 

$$ r = \{ [x_1, \dots, x_n] \mapsto [y_1, \dots, y_m] :  c_1, \dots, c_p \}$$

The read and write access functions of a loop nest can be modeled with relations. The read relations in the Figure \ref{matmulcode} code are shown below.

\begin{align*}
r_1  = \{ S[i, j, k] \mapsto  C[i, j] \} \\
r_2  = \{ S[i, j, k] \mapsto  A[i, k] \}\\
r_3  = \{ S[i, j, k] \mapsto  B[k, j] \}
\end{align*}

The sole write relation in the loop is:\\ $w_1 = S[i, j, k] \mapsto C[i, j]$.

\paragraph{Apply operation} When a relation $r$ is applied on a set $s$, the domain of $r$ will be intersected with $s$ and the resulting range will be a new set $s'$. The set $s'$ is said to be the result of the apply operation. The operation is mathematically defined as:

$$ ( \vec{y} \in s') \Longleftrightarrow (\exists \vec{x} ~~\text{s.t}~~ (\vec{x} \in s \land \vec{x} \mapsto \vec{y}) \in r )$$

The data footprint of the loop can be computed by \emph{applying} read and write \emph{relations} on the iteration space \emph{set}:

$$ r_1(I) \cup r_2(I) \cup r_3(I) \cup w_1(I) $$

\subsection{Polyhedral dependences}
The exact data dependences in loop nests can be computed in the polyhedral model and are expressed as maps from source iterations to target iterations involved in the dependence. For cache data reuse analysis developed in \S\ref{sec:reuse}, we consider  four kinds of dependences -- Read-After-Read (RAR), Read-After-Write (RAW, a.k.a \emph{flow}), Write-After-Read (WAR, a.k.a \emph{anti}), and Write-After-Write (WAW). 
  The data dependencies of the matrix multiplication code in Figure \ref{matmulcode} are shown below.
  
\begin{align*}
d_1 = & \{ S[i, j, k] \mapsto S[i', j', k'] : i' = i \land j' = j \land k < k' < K \} \\
d_2 = & \{ S[i, j, k] \mapsto S[i', j', k'] : i' = i \land k' = k \land j < j' < N \} \\
d_3 = & \{ S[i, j, k] \mapsto S[i', j', k'] : j' = j \land k' = k \land  i < i' < M \} \\
\end{align*}

The dependence $d_2$ is induced by array reference A[i][k]. An element of array A, say A[0][0] which is accessed in \emph{source} iteration $[i=0,j=0,k=0]$ gets reused in \emph{target} iterations $[i'=0,j'>0,k'=0]$. The source to target iteration relationships such as this are expressed in a parametric fashion as the relation $d_2$. 

\section{Cache Data Reuse Analysis}
\label{sec:reuse}

We develop a polyhedral model based cache data reuse analysis to characterize a loop-nest's behavior with respect to a given cache hierarchy. The analysis computes the various existing data reuses of a program and then for the input cache hierarchy determines which data reuses are exploitable at various levels of cache.

\subsection{Working set size computation}
\label{sec:wscompute}

Each data dependence in a loop is also a case of data reuse -- the source and target iterations involved in the dependence touch the same data element and therefore, the data is reused. For a data dependence and hence data reuse to be realizable in a given level of cache, all the data elements accessed between the source and target iterations of the dependence -- the \emph{working set} -- have to be retained in the cache so that when the execution reaches the target iteration, the data element(s) used in the source iteration will still be present in the cache. 

\begin{algorithm}  
 \KwInput{Loop nest: $\mathcal{L}$}
 \KwOutput{The working set sizes: $WS_{all}$}
\{Iteration space: $\mathcal{I}$, Read relations: $r_{read}$, Write relations: $r_{write}$, Schedule: $\delta$\} $\leftarrow$ Parse the loop nest $\mathcal{L}$ \\
\{$\mathcal{D}_{RAR}, \mathcal{D}_{RAW}, \mathcal{D}_{WAR}, \mathcal{D}_{WAW}$\} $\leftarrow$ Compute read-after-read, read-after-write, write-after-read, write-after-write dependences of $\mathcal{L}$ \\
$\mathcal{D}_{all} \leftarrow \mathcal{D}_{RAR} \cup \mathcal{D}_{RAW} \cup \mathcal{D}_{WAR} \cup \mathcal{D}_{WAW}$ \\
$WS_{all} \leftarrow \emptyset$ \\
\tcc{Iterate through all dependences to compute the working set sizes}
\For{$d \in \mathcal{D}_{all}$}{
 $\mathcal{I}_{source} \leftarrow \text{lexmin dom}~~~ d$ \\
 $\mathcal{I}_{min\_tar} \leftarrow \text{lexmin}~~~ d(\mathcal{I}_{source})$ \\
 $\mathcal{I}_{max\_tar} \leftarrow \text{lexmax}~~~ d(\mathcal{I}_{source})$ \\
 $\mathcal{I}_{min\_WS} \leftarrow (\mathcal{I} <<= \mathcal{I}_{min\_tar}) 
 -  (\mathcal{I} << \mathcal{I}_{source})$\\
  $\mathcal{I}_{max\_WS} \leftarrow (\mathcal{I} <<= \mathcal{I}_{max\_tar}) 
 -  (\mathcal{I} << \mathcal{I}_{source})$\\
 $WS_{min} \leftarrow | r_{read}(\mathcal{I}_{min\_WS}) \cup r_{write}(\mathcal{I}_{min\_WS})|$ \\
  $WS_{max} \leftarrow | r_{read}(\mathcal{I}_{max\_WS}) \cup r_{write}(\mathcal{I}_{max\_WS})|$ \\
  Add $WS_{min}$ and $WS_{max}$ to $WS_{all}$
}
\caption{Compute working set sizes}
\label{alg:wscompute}
\end{algorithm}

Algorithm \ref{alg:wscompute} computes all the working sets of the input loop nest.
First, the input C source file is parsed using the Polyhedral Extraction Tool (PET) \cite{Verdoolaege2012pet} to obtain the polyhedral representation of the program, namely iteration space of the loop nest, read and write relations and the schedule (line 1).  The exact (and not transitive) RAR, RAW, WAR, WAW dependences are then computed and a union of all the four kinds of dependences is formed (line 2 and 3). 
The task now is to compute the working set size for each dependence which is carried out from line 5 through 13. For a given dependence, we consider a representative \emph{source} -- the first iteration (lexicographically) of all the source iterations (line 6).
We can now compute the target iterations for the lexicographically first/minimum iteration. If the data element that is used in the source iteration is used in multiple subsequent iterations then there may be multiple target iterations for the same source iteration. Therefore, the working sets to exploit the data reuse may vary. 
For this reason, we compute the first (or lexicographically minimum) and last (lexicographically maximum) iterations of the target iterations (line 7 and 8).
The intervening iterations between source and the first target iteration are determined (line 9). Similarly, the iterations between source and the target iteration are derived (line 10).
The working sets will be the union of all the read and written data elements between 
the source and the first/last iterations of the target iteration set (line 11 and 12).
Correspondingly, for a dependence we compute two working set sizes -- $WS_{min}$ and $WS_{max}$, if there are multiple target iterations for a source iteration in a given dependence.
What this means is, in order to be able to exploit at least one data reuse arising from the dependence $d$, the cache memory should be capacious enough to hold at least $WS_{min}$ data elements. If all the data reuses are to be realized -- till the last target iteration, then the cache should of size equal to $WS_{max}$ in terms of the datatype's elements.

We illustrate the operation of the algorithm using the running example in Figure \ref{matmulcode}. Let us examine the dependence:
$$d_2 =  \{ S[i, j, k] \mapsto S[i', j', k'] : i' = i \land k' = k \land j < j' < N \}$$
Of all the source iterations, the first/lexicographically minimum iteration is: 
$$\mathcal{I}_{source} = \{ S[i = 0, j = 0, k = 0] \}$$
Its target iterations are: 
$\{ S[i = 0, j, k = 0] :  0 < j < N \}$
Among the target iterations, the first one is: 
$$I_{min\_tar} = \{ S[i = 0, j = 1, k = 0]  \}$$ 
and the last one is: 
$$I_{max\_tar} = \{ S_3[i = 0, j = N-1, k = 0]  \}$$

The number of data elements of the three arrays -- A, B, C accessed between $\mathcal{I}_{source}$ and $I_{min\_tar}$ is derived by \emph{applying} the read and write relations on the intervening iteration set and it is: $$WS_{min} = 2K + 3$$

The $K$ elements of array A -- $A[0][0, 1, \dots, K-1]$, the $K+1$ elements of array B --
$B[0, 1, \dots, K-1][0]$ and $B[0][1]$, and finally $2$ elements of array C -- 
$C[0][0], C[0][1]$ accessed between the source iteration $S[i = 0, j = 0, k = 0]$
and the target iteration $I_{min\_tar} = S[i = 0, j = 1, k = 0]$ lead to the $WS_{min}$ size of $2K + 3$.

The maximum working set size -- the size of the data touched between $\mathcal{I}_{source}$ and $I_{max\_tar}$ is:

$$ WS_{max} = N\times K + N +1 $$
The $WS_{max}$ size is arrived at by counting the number of array elements
accessed between the source iteration - $S[i = 0, j = 0, k = 0]$ and the target iteration - 
$I_{max\_tar} = \{ S_3[i = 0, j = N-1, k = 0]  \}$.
As far as array A is concerned, $K$ elements of it -- $A[0][0, 1, \dots, K-1]$ are read.
Array B's elements -- $B[0, 1, \dots, K-1][0, 1, \dots, N-2]$ plus $B[0][N-1]$ are read which total $K \times (N-1) + 1$.
$N$ elements of array C are read and written -- $C[0][0, 1, \dots, N-1]$. Therefore, a total of $N\times K + N +1$ are read and written.

\subsection{Poly-ranking algorithm}
\label{sec:wholepolyranking}

\begin{algorithm} 
 \KwInput{The working set sizes: $WS_{all}$, \\
  Cache sizes: $Size_{L_1}, \dots Size_{L_n}$} 
 \KwOutput{Working set sizes per cache: $WS^{L_i} ~~\text{for}~~ i = 1, \dots, n$, \\
 Memory working set size: $WS^{mem}$}

Initialize $WS^{L_i} ~~\text{to} ~~ 0~~\text{for}~~ i = 1, \dots, n$, \\
Sort working set sizes in $WS_{all}$ from smallest to largest \\
\For{$WS_j \in WS_{all}$}{
	\For{$Size_{L_i} \in Size_{L_1}, \dots Size_{L_n}$}{
		\If{$(WS_j + WS^{L_i}) \le Size_{L_i}$} {
			$WS^{L_i} = WS^{L_i} + WS_j$ \\
			\textbf{break} \\
		}
	}
}

Add the working sets $WS_j \in WS_{all}$ that do not fit any cache to $WS^{mem}$
\caption{Compute working set sizes w.r.t cache sizes}
\label{alg:cachewscompute}
\end{algorithm}

\begin{figure}[h!]
\centering
\includegraphics[scale=0.5]{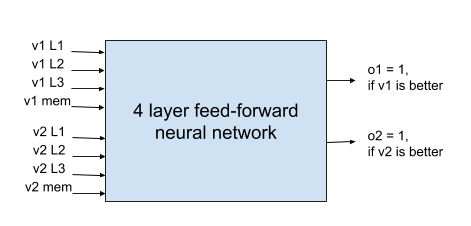}
\caption{The DNN architecture for ranking of code variants}
\label{fig:polydnn}
\end{figure}

We have built a code generator to emit a number of program variants.
The code generator creates the loop variants by applying tiling and loop interchange
program transformations. The tile sizes are varied as well.
The working set size computation analysis --\S\ref{sec:wscompute} is performed on each program version generated.
Among the many variants generated, the poly-ranking algorithm described below picks
the top $k$ best performing versions, where $k$ is a parameter.
In this work, we pick the top 1 variant
the code generator produces, i.e., $k = 1$.

If the working set size corresponding to a data reuse in the program is smaller
than the cache size then the data reuse is exploitable in the cache.
The poly-ranking system considers caches at different levels (typically L1, L2, and L3)
and for each data reuse, determines at what level of cache hierarchy is the data reuse
realizable.
Algorithm \ref{alg:cachewscompute} shows the steps to determine the cumulative
working set sizes at each level of cache. The inputs to the algorithm are the 
working set sizes computed for a loop nest, and the cache sizes of the target system.
The algorithm determines the fastest level of cache where the working set size corresponding to each data reuse fits
and adds it to that cache's working set size. 
The working set sizes that fit in a particular level of cache $L_i$ are denoted by $WS^{L_i}$.
If a working set does not fit in any cache, then the data reuse happens
out of the main memory. Consequently, the memory's working set size is updated.

\subsubsection{Performance cost model based ranking}
\label{sec:polyranking}
The running time of the loop is directly related to the latency of the cache where the data
reuse occurs as well as the working set size. Furthermore, the running time is inversely
related to the bandwidth of the cache.
Based on these observations, we define the following cost function:

\begin{align}
\mathcal{C} = & \sum_{L_i} WS^{L_i} \times \frac{lat^{L_i}}{bw^{L_i}}  + WS^{mem} \times \frac{lat^{mem}}{bw^{mem}} \label{eq1} 
\end{align}

The latency of cache $L_i$ is $lat^{L_i}$ while its bandwidth 
is denoted by $bw^{L_i}$.
For each code variant generated, we run the cache data reuse analysis and
calculate the above cost function. 
Then, the variants are ranked in the decreasing order of the value of the cost function:
the lower the value, the higher is its assumed performance, and higher is its rank.

\subsubsection{DNN-based ranking algorithm}
\label{sec:dnnranking}

We explore the use of deep neural networks (DNNs) for 
ranking code variants. 
For the purposes of training the DNN model, 
we collect the performance data of code variants generated and
the statistics as outputted by Algorithm \ref{alg:cachewscompute} -- 
working set sizes at different levels of the memory hierarchy.

We train the DNN model to perform relative ordering of \emph{two} code
variants.
We then use a \emph{tournament} based ranking system to assign ranks
to the different code versions created -- 
we play each code variant against every other code variant.
For each variant, we record the number of wins it has accumulated.
We then rank the variants based on the number of wins -- 
the higher the number of wins, the higher the rank.

Figure \ref{fig:polydnn} shows the architecture of the DNN.
We normalize the compiler generated statistics of two code variants 
in the following fashion and input them to the DNN.
We sum the working set sizes of the two variants together:
$sum = v_1~ L_1 + v_1~ L_2 + v_1~ L_3 + v_1~ mem + v_2~ L_1 + v_2~ L_2 + v_2~ L_3 + v_2~ mem$
and divide the individual statistic by this sum.
The rationale for considering the sum of the two statistics together is that
if one of the variants is creating higher volume working set sizes then
its statistics should appear bigger to the DNN.
This is because the smaller the working set sizes, we can expect higher
performance.
Therefore, for the DNN to learn the relative performances of the two variants,
it is crucial that it sees the relative sizes of the working set sizes.
Normalizing each variant individually (by considering the sum of statistics of one
variant alone) would not bring out the differences in the absolute values
of working set sizes of the two variants at different cache levels.

We use four intermediate layers of 64, 32, 16, 8 neurons respectively.
We use \emph{relu}, \emph{relu}, \emph{softsign}, and \emph{relu}
activation functions for the four intermediate layers.
The output layer consists of two neurons and we use the \emph{softmax}
function for the output layer.
The values of the two output neurons, because of the use of softmax function,
sum to 1. 
If the output value is above a threshold - $\theta$, we consider it a 1, otherwise a 0.
If the first neuron fires a 1, then the first variant is considered the winner.
If the second neuron fires a 1, then the second variant is considered the winner.
If both of them are zero because none of them are above the threshold, then
it is a draw between the two variants. In this work, we set the threshold $\theta$ to 0.6.

We experimented with deeper models as well. However, the depth beyond 
four layers did not have any discernible effect on accuracy.

\begin{figure*}[h!]
\begin{lstlisting}[language=C,basicstyle=\normalsize,frame=bottomline]
#pragma omp parallel for private(ofm_tile, ifm_tile, ij, oj, kj, ki, ii)
for (img = 0; img < nImg; ++img) {
 for (ofm_tile = 0; ofm_tile < nOfm / GEMM_BLOCK; ++ofm_tile) {
  for (ifm_tile = 0; ifm_tile < nIfm / GEMM_BLOCK; ++ifm_tile) {
   for (oj = 0; oj < ofh; ++oj) {
	ij = oj * STRIDE_H;
	for (kj = 0; kj < kh; ++kj) {
	 for (ki = 0; ki < kw; ++ki) {

	  /* GEMM operation begins */
	  for (oi = 0; oi < ofw; ++oi) {
	   ii = oi * STRIDE_W;
	    for (ofm = 0; ofm < GEMM_BLOCK; ++ofm) {
		for (ifm = 0; ifm < GEMM_BLOCK; ++ifm) {
		 output[img][ofm_tile][oj][oi][ofm] +=
		  filter[ofm_tile][ifm_tile][kj][ki][ifm][ofm] 
		   * input[img][ifm_tile][ij+kj][ii+ki][ifm];
		}
	   }
	  }
	 /* GEMM operation ends */
     }
    }
   }
  }
 }
}
\end{lstlisting}
\caption{The 2-D Convolution  code}
\label{fig:convcode}
\end{figure*}


\section{Experimental Evaluation}
\label{sec:experiments}

We conduct experiments to evaluate the efficacy of the PolyScientist system which combines
loop optimizations with expert-coded kernels.
Since the aim of PolyScientist is to achieve performance competitive with manually optimized code and at the same time retaining the attractiveness of an automatic
compiler, 
we gauge the performance of PolyScientist against a state-of-the-art library created
specifically for deep learning networks -- the latest version of Intel MKL-DNN \cite{intelmkldnn} viz., v1.0.4 and 
against the compiler generated code
using the Intel icc compiler of version 19.0.3.199.

\subsection{Set up}
We use the PolyScientist system to optimize the convolutions of Resnet-50 \cite{he2016deep}, Fast R-CNN (\textsf{fastrcnn}) \cite{girshick2015fast},  Mask R-CNN (\textsf{maskrcnn})  \cite{DBLP:journals/corr/HeGDG17}, Xception (\textsf{xception}) \cite{DBLP:journals/corr/Chollet16a}, You Only Look Once v2 (\textsf{yolov2}) \cite{DBLP:journals/corr/RedmonDGF15}, 
MobileNets (\textsf{mobilenet}) \cite{DBLP:journals/corr/HowardZCKWWAA17},
AlexNet (\textsf{alexnet}) \cite{Krizhevsky:2012:ICD:2999134.2999257},
OverFeat (\textsf{overfeat}) \cite{sermanet2013overfeat}
GoogLeNet v1 and v3 \cite{43022},
and 
(\textsf{googlenetv1}, \textsf{googlenetv3}),
 the popular
and the state-of-the-art image recognition neural network models.
We also gather the performance results from the hand-coded, highly optimized Intel MKL-DNN library for the same convolutions.
In this work,
we evaluate the performance benefits of the PolyScientist system on CPUs.
 Since in today's datacenters, CPUs are predominantly used for inference tasks
partly due to latency considerations, 
we study the performance of forward-pass convolutions
which are used in the \emph{inference} tasks while performing image recognition.

Figure \ref{fig:convcode} shows the convolution code
with the \emph{GEMM} (matrix multiplication) microkernel replaced with the equivalent C code.
The shown code is data tiled in the input and output channel 
dimensions.
The convolution code has a matrix multiplication operation
(denoted \emph{GEMM} in the code) embedded in it.
The loops corresponding to matrix multiplication
are moved to the inner most positions so that matrix multiplication is performed
on the fastest varying dimensions of the arrays. 
We use the performance obtained using the code
shown in \ref{fig:convcode} as the \emph{baseline}.

We use the LIBXSMM \cite{libxsmm}
implementation for matrix multiplication --
the \emph{microkernel}.
\emph{PolyScientist} performs outer loop optimization around the call to the matrix multiplication microkernel
by loop reordering and tiling using various tile sizes.
We show the performance obtained by inserting
the LIBXSMM microkernel in the code listed in Figure \ref{fig:convcode}
under the banner of \textsf{Microkernel} in the subsequent performance
graphs.
Comparing the performance of \textsf{Microkernel} with 
\textsf{PolyScientist} will show the need to perform outer loop tuning
as done by \textsf{PolyScientist} to obtain high performance for all layers
and for all models.

\begin{figure*}[h!]
\centering
\begin{minipage}{0.49\textwidth}
\centering
\includegraphics[scale=0.5]{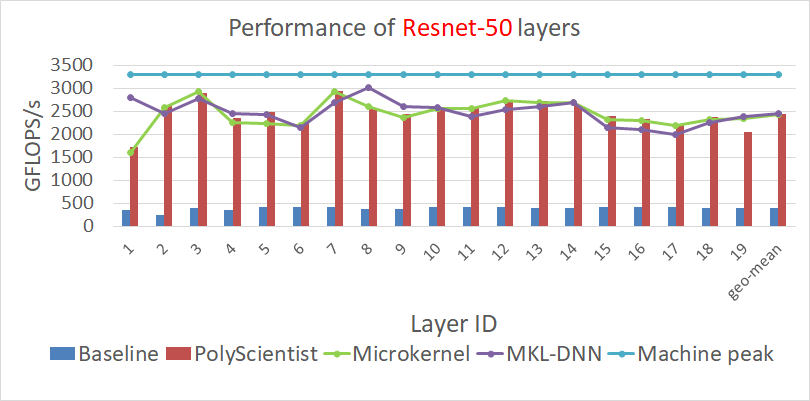}
\caption{Performance of Resnet-50 layers on a 28-core Intel Cascade Lake server}
\label{fig:resnet}
\end{minipage}
\begin{minipage}{0.49\textwidth}
\centering
\includegraphics[scale=0.5]{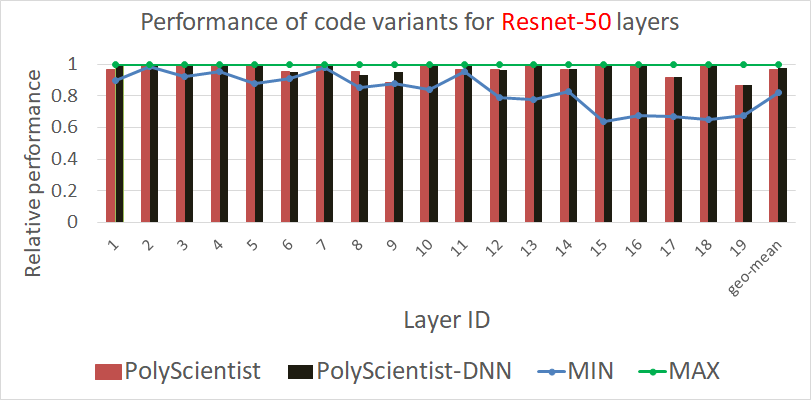}
\caption{Performance distribution of code variants for Resnet-50 layers on a 28-core Intel Cascade Lake server}
\label{fig:resnet_distro}
\end{minipage}
\end{figure*}

\begin{figure*}[h!]
\centering
\begin{minipage}{0.49\textwidth}
\centering
\includegraphics[scale=0.5]{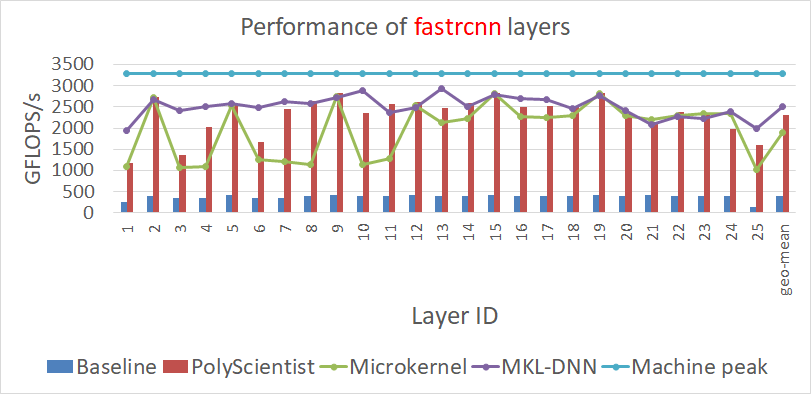}
\caption{Performance of fastrcnn layers on a 28-core Intel Cascade Lake server}
\label{fig:fastrcnn}
\end{minipage}
\begin{minipage}{0.49\textwidth}
\centering
\includegraphics[scale=0.5]{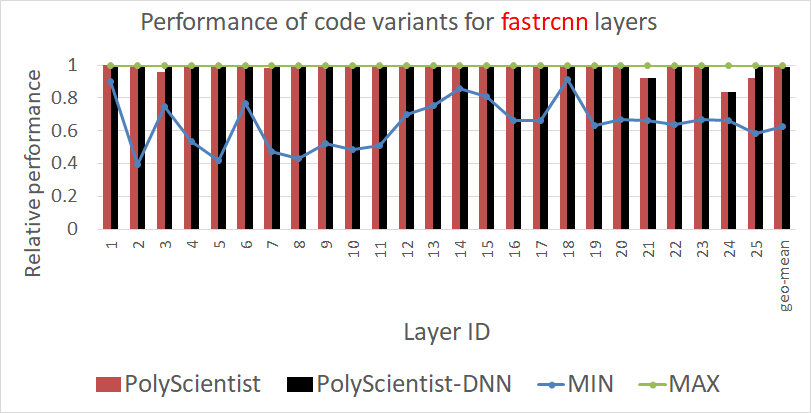}
\caption{Performance distribution of code variants for fastrcnn layers on a 28-core Intel Cascade Lake server}
\label{fig:fastrcnn_distro}
\end{minipage}
\end{figure*}

\begin{figure*}[h!]
\centering
\begin{minipage}{0.49\textwidth}
\centering
\includegraphics[scale=0.5]{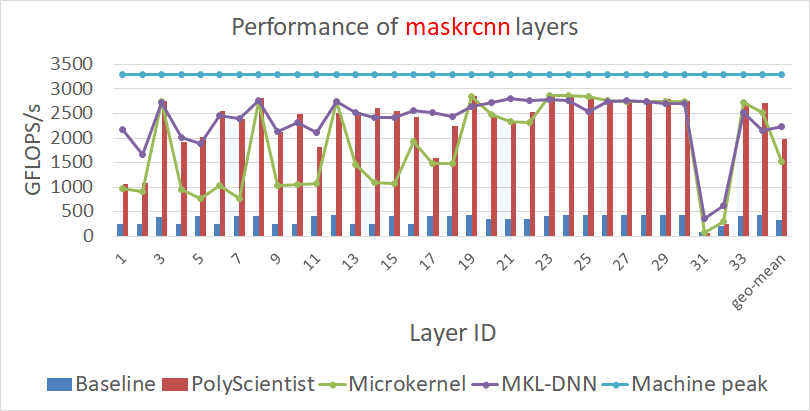}
\caption{Performance of maskrcnn layers on a 28-core Intel Cascade Lake server}
\label{fig:maskrcnn}
\end{minipage}
\begin{minipage}{0.49\textwidth}
\centering
\includegraphics[scale=0.5]{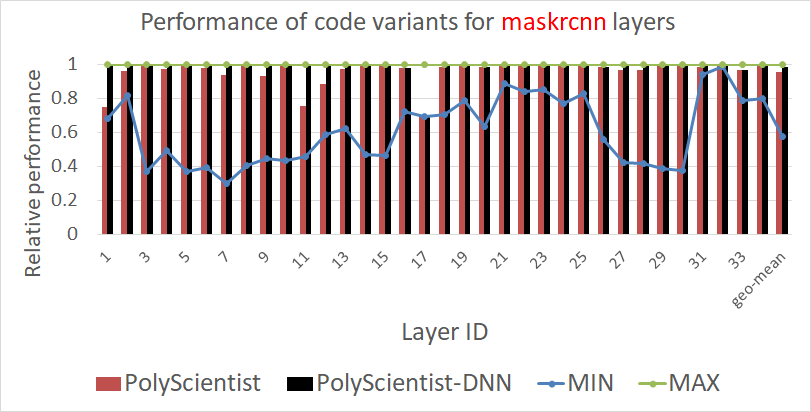}
\caption{Performance distribution of code variants for maskrcnn layers on a 28-core Intel Cascade Lake server}
\label{fig:maskrcnn_distro}
\end{minipage}
\end{figure*}

\begin{figure*}[h!]
\centering
\begin{minipage}{0.49\textwidth}
\centering
\includegraphics[scale=0.5]{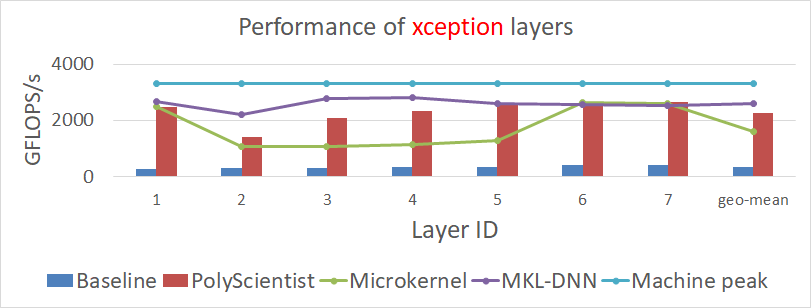}
\caption{Performance of xception layers on a 28-core Intel Cascade Lake server}
\label{fig:xception}
\end{minipage}
\begin{minipage}{0.49\textwidth}
\centering
\includegraphics[scale=0.5]{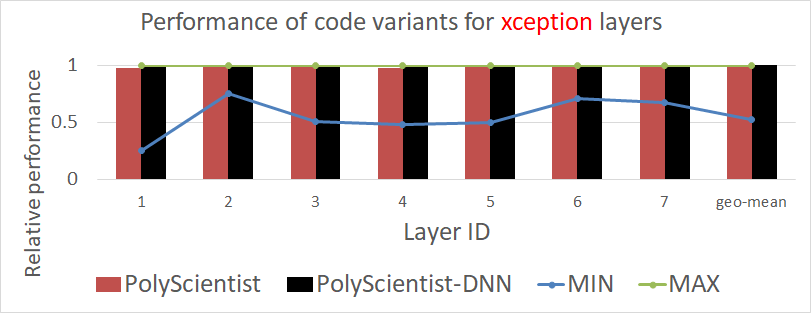}
\caption{Performance distribution of code variants for xception layers on a 28-core Intel Cascade Lake server}
\label{fig:xception_distro}
\end{minipage}
\end{figure*}

\begin{figure*}[h!]
\centering
\begin{minipage}{0.49\textwidth}
\centering
\includegraphics[scale=0.5]{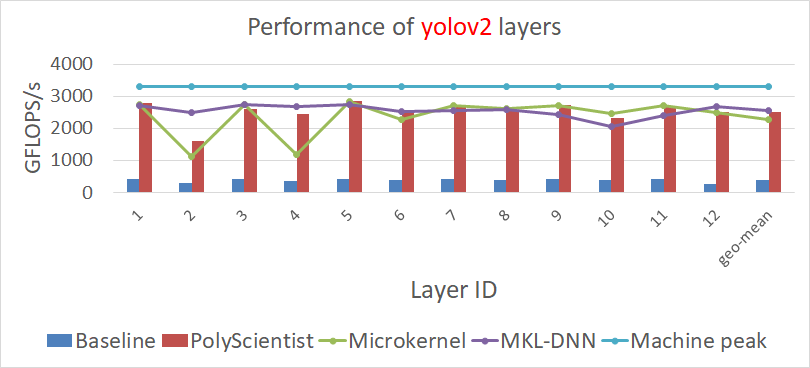}
\caption{Performance of yolov2 layers on a 28-core Intel Cascade Lake server}
\label{fig:yolov2}
\end{minipage}
\begin{minipage}{0.49\textwidth}
\centering
\includegraphics[scale=0.5]{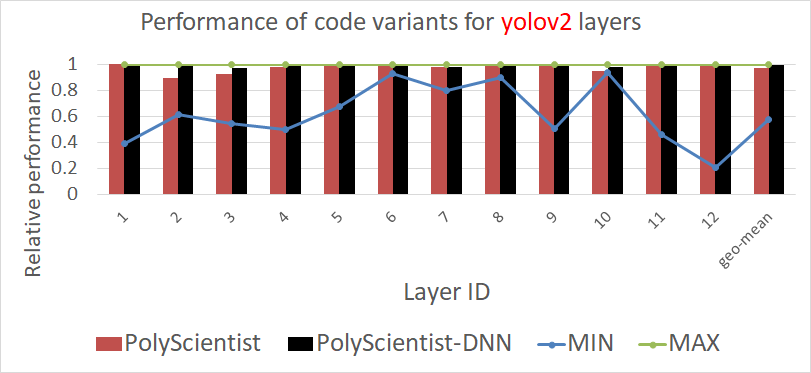}
\caption{Performance distribution of code variants for yolov2 layers on a 28-core Intel Cascade Lake server}
\label{fig:yolov2_distro}
\end{minipage}
\end{figure*}

\begin{figure*}[h!]
\centering
\begin{minipage}{0.49\textwidth}
\centering
\includegraphics[scale=0.5]{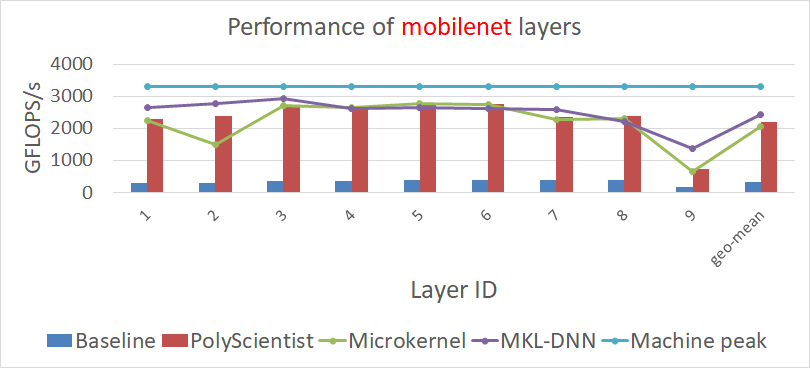}
\caption{Performance of mobilenet layers on a 28-core Intel Cascade Lake server}
\label{fig:mobilenet}
\end{minipage}
\begin{minipage}{0.49\textwidth}
\centering
\includegraphics[scale=0.5]{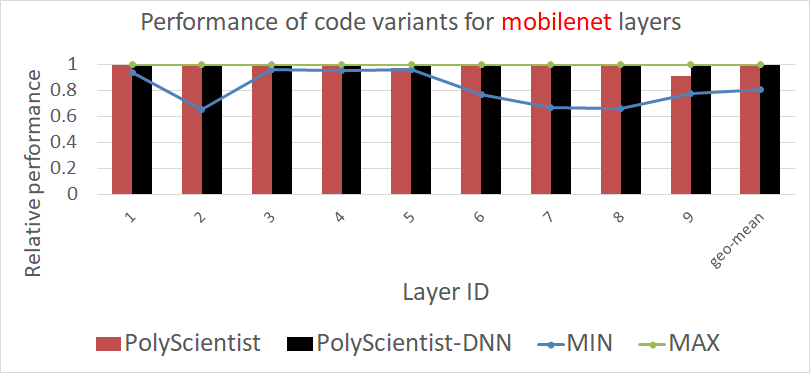}
\caption{Performance distribution of code variants for mobilenet layers on a 28-core Intel Cascade Lake server}
\label{fig:mobilenet_distro}
\end{minipage}
\end{figure*}

\begin{figure*}[h!]
\centering
\begin{minipage}{0.49\textwidth}
\centering
\includegraphics[scale=0.5]{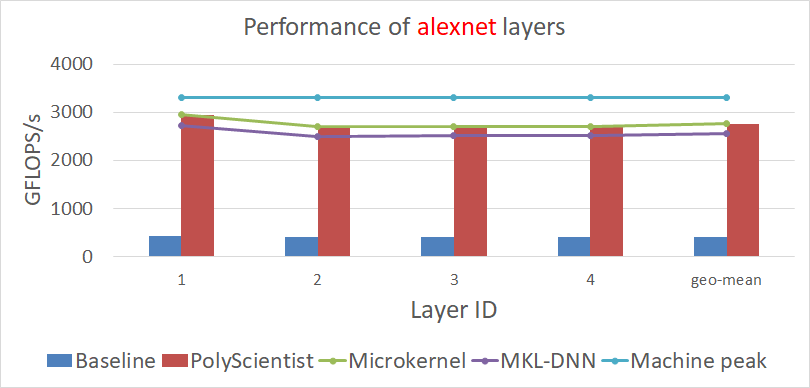}
\caption{Performance of alexnet layers on a 28-core Intel Cascade Lake server}
\label{fig:alexnet}
\end{minipage}
\begin{minipage}{0.49\textwidth}
\centering
\includegraphics[scale=0.5]{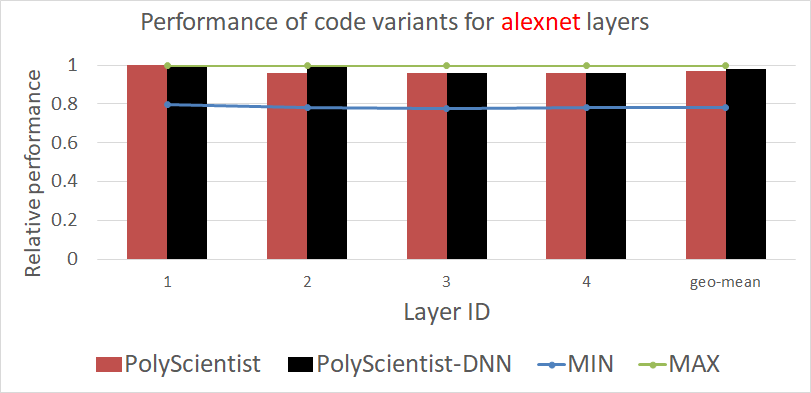}
\caption{Performance distribution of code variants for alexnet layers on a 28-core Intel Cascade Lake server}
\label{fig:alexnet_distro}
\end{minipage}
\end{figure*}

\begin{figure*}[h!]
\centering
\begin{minipage}{0.49\textwidth}
\centering
\includegraphics[scale=0.5]{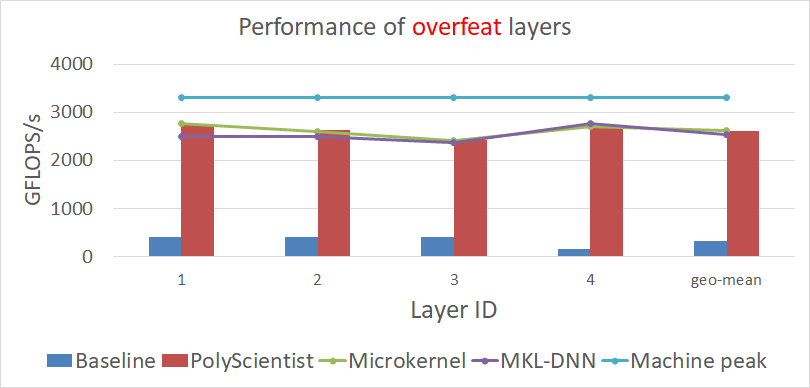}
\caption{Performance of overfeat layers on a 28-core Intel Cascade Lake server}
\label{fig:overfeat}
\end{minipage}
\begin{minipage}{0.49\textwidth}
\centering
\includegraphics[scale=0.5]{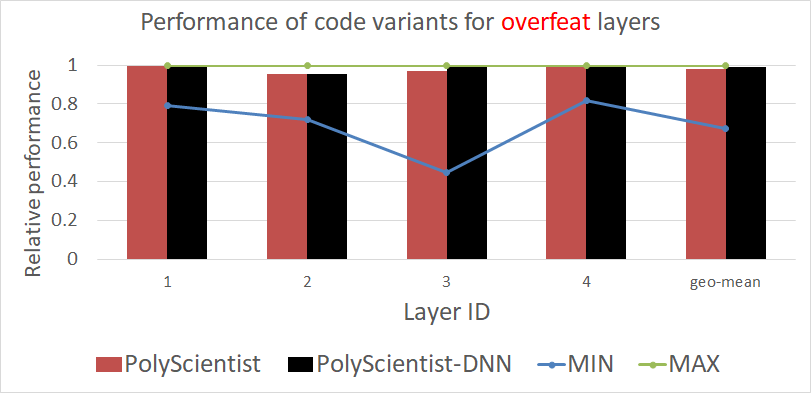}
\caption{Performance distribution of code variants for overfeat layers on a 28-core Intel Cascade Lake server}
\label{fig:overfeat_distro}
\end{minipage}
\end{figure*}

\begin{figure*}[h!]
\centering
\begin{minipage}{0.49\textwidth}
\centering
\includegraphics[scale=0.5]{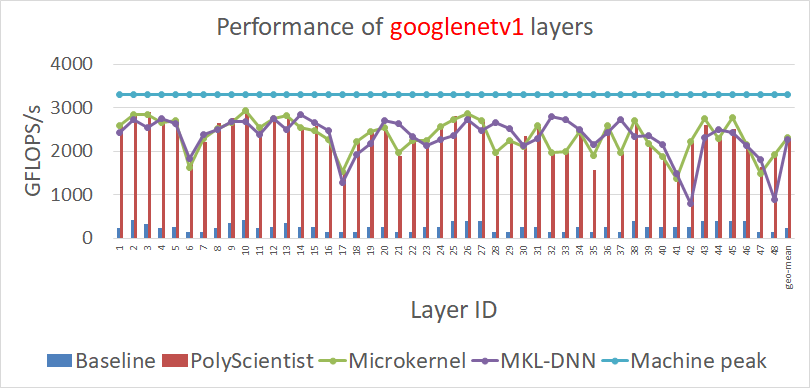}
\caption{Performance of googlenetv1 layers on a 28-core Intel Cascade Lake server}
\label{fig:googlenetv1}
\end{minipage}
\begin{minipage}{0.49\textwidth}
\centering
\includegraphics[scale=0.5]{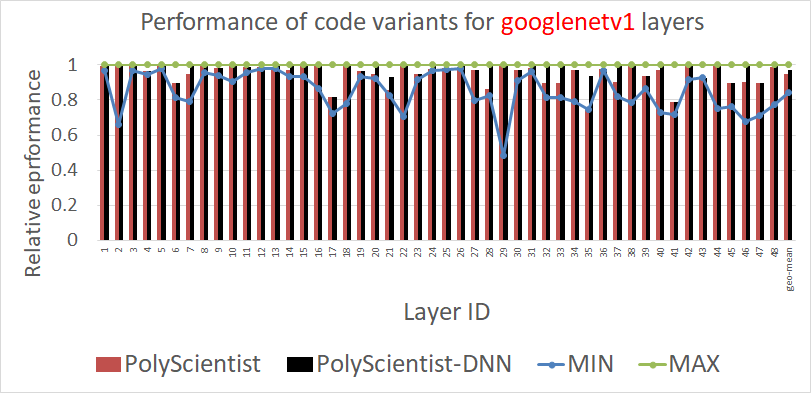}
\caption{Performance distribution of code variants for googlenetv1 layers on a 28-core Intel Cascade Lake server}
\label{fig:googlenetv1_distro}
\end{minipage}
\end{figure*}

The experiments are run on the latest Intel servers -- Intel(R) Xeon(R) Platinum 8280 (Cascade Lake) CPU servers running at 
the frequency of 2.70GHz.
Each processor has 28 cores, 32KB private L1 cache, 1MB private L2 cache, and
39MB shared L3 cache.
Each program is run a 1000 times and the average performance across those runs
is reported in the paper.
The machine has a 512-bit SIMD vector unit and supports
AVX-512 vector instructions. 
Consequently, 16 floating point arithmetic operations can be performed
at a time (each floating point number is 32 bits long, and therefore,
16 floating point numbers make up 512 bits: $32 \times 16 = 512$).
Since the microkernel vectorizes along the input and 
output channel loops ($ifm$ and $ofm$ loops in the code),
to fully utilize the vector unit, the input and output channel
widths have to be 16 or multiples of 16.
In the CNN models considered, $86\%$ of the convolutions meet
this criterion and those convolutions are selected for experimental
evaluation.
The peak single precision floating point performance of a
Cascade Lake processor is  \textasciitilde 3,300 GFLOPS/s.
We set the mini-batch size to 28 and use data parallelism:
the convolution operator is applied on 28 images simultaneously.

To train a DNN model for performing ranking of code variants as described
in \S \ref{sec:dnnranking}, we use 70\% of the experimental data collected
(to avoid overfitting).
We create a single DNN model using data from all CNN models
and use it to rank variants across the CNN models.

\begin{figure*}[h!]
\centering
\begin{minipage}{0.49\textwidth}
\centering
\includegraphics[scale=0.5]{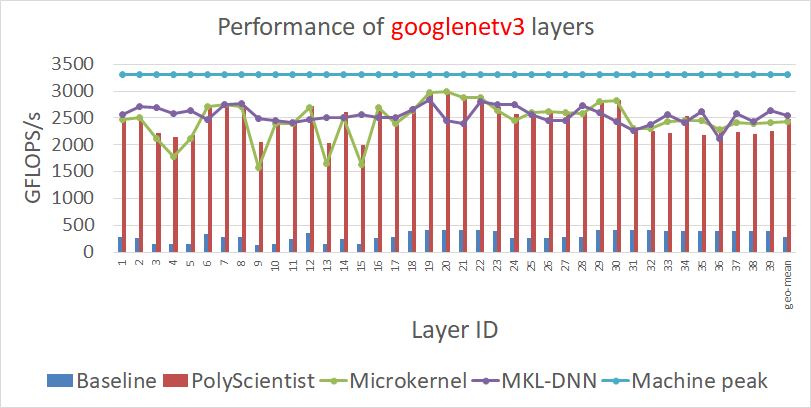}
\caption{Performance of googlenetv3 layers on a 28-core Intel Cascade Lake server}
\label{fig:googlenetv3}
\end{minipage}
\begin{minipage}{0.49\textwidth}
\centering
\includegraphics[scale=0.5]{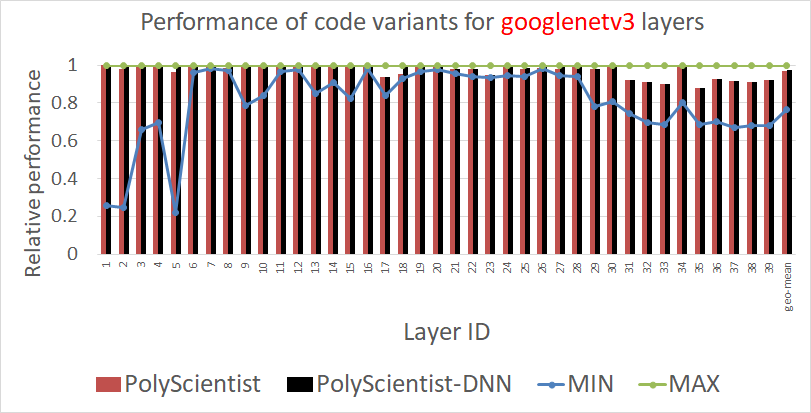}
\caption{Performance distribution of code variants for googlenetv3 layers on a 28-core Intel Cascade Lake server}
\label{fig:googlenetv3_distro}
\end{minipage}
\end{figure*}


\subsection{Experimental Results}

\begin{table}
\caption{The geometric average of the performance of layers of various models in GFLOPS/s on a 28-core Intel Cascade Lake server}
\label{table:perf}
\small
\begin{center}
 \begin{tabular}{|c | r | r | r |  r|} 
 \hline
Model & {\scriptsize Baseline} & {\scriptsize	PolyScientist} & {\scriptsize	PolySct.-DNN} & {\scriptsize MKL-DNN} \\ [0.5ex] 
 \hline\hline
Resnet-50 & 391 & 2448 & 2463 & 2459 \\ \hline
fastrcnn & 390 & 2305 & 2312 &  2511 \\  \hline
maskrcnn & 336 & 1988 & 2052 & 2243 \\ \hline
xception & 348 & 2276 & 2292 & 2595 \\ \hline
yolov2 & 387 & 2503 & 2551 & 2547 \\ \hline
mobilenet & 333 & 2217 & 2247 &  2440 \\ \hline
alexnet & 418 & 2751 & 2779 & 2563 \\ \hline
overfeat & 331 & 2621 & 2644 &  2534 \\ \hline
googlenetv1 & 241 & 2322 & 2377 & 2276 \\ \hline
googlenetv3 & 291 & 2480 & 2492 & 2547 \\ \hline
 
 \hline
\end{tabular}
\end{center}
\end{table}

Figure \ref{fig:resnet} shows the performance in terms of GFLOPS/s
(Giga Floating point Operations per second) of 
the baseline code, PolyScientist, Microkernel and MKL-DNN 
on convolutions of Resnet-50. 
The PolyScientist performance shown is the performance of 
the top code variant selected using the cost modeling based
poly-ranking algorithm described in \S \ref{sec:polyranking}.
The performance of PolyScientist over the baseline is 
5X to 10X for all layers.
The higher performance of PolyScientist is due to 
1)  the use of optimized GEMM microkernel and
2)  the optimization of outer loops around the call to the
microkernel.
In most cases, PolyScientist closely matches the performance of
MKL-DNN library.
In several instances, PolyScientist outperforms MKL-DNN, 
notably for layers with IDs 15, 16, and 17 where
the performance gain is 11\%.
On some layers such as layer 1, MKL-DNN fares better.
This is explained by customizations for specific problem sizes including
insertion of careful data prefetching instructions in the
 MKL-DNN library code.
In contrast, PolyScientist's approach is automatic and 
in the case of Resnet-50, we observe that we are able to attain
the same performance levels as MKL-DNN.
The geometric average of GFLOPS/s numbers are also shown
in the graph.
We report the geometric average of performance results in 
Table \ref{table:perf} as well for Resnet-50 and all other CNN models.
For Resnet-50, we observe that performance of \textsf{Microkernel} and \textsf{PolyScientist} is similar indicating that the original loop order shown in 
Figure \ref{fig:convcode} gets good performance.

Figure \ref{fig:resnet_distro} shows the performance 
distribution of code variants generated for each layer of Resnet-50.
The performance is normalized with respect to that of the best
performing variant found empirically.
The crux of the PolyScientist technology presented in the paper
is to rank a given set of code variants 
using compile-time static analysis.
Therefore, the closer the performance of the PolyScientist picked
version is to the maximum performance seen by any code 
variant explored, the more efficacious the PolyScientist algorithms are.
In the graph, we show the minimum performance observed,
the maximum performance seen, the performance of the code
picked per the poly-ranking algorithm (\S \ref{sec:polyranking})
-- \textsf{PolyScientist}
and the performance of  the code picked per the DNN based ranking algorithm
(\S \ref{sec:dnnranking}) -- \textsf{PolyScientist-DNN}.
We note that the performance of \textsf{PolyScientist} version
is close to the maximum performance in most layers save layer 19.
Even though in terms of cache behavior (PolyScientist primarily models the cache
behavior), the variant selected by PolyScientist may be the best, other factors such as
prefetching, TLB behavior etc may cause their performance to be lower than those
of other variants. 
The minimum performance seen i.e., the performance of the worst 
code variant, varies across layers -- for layer 12 through 19,
the minimum performance is much farther from the maximum
performance. For the initial layers however, the different
code variants generated perform similarly.
In all cases, we note that \textsf{PolyScientist}
performs significantly better than the worst variant including
layer 19 where \textsf{PolyScientist} picked code is 14\%
slower than the best performing version and is 28\%
higher performing than the worst variant.
We observe that there is no considerable difference
in the performance achieved by 
cost model based ranking method -- \textsf{PolyScientist}, 
and the DNN
based ranking method -- \textsf{PolyScientist-DNN}.

The performance achieved by different methods for convolutions of 
\textsf{fastrcnn} is shown in Figure \ref{fig:fastrcnn}.
The performance of \textsf{PolyScientist} vis-a-vis the baseline code is 
anywhere between 4X and 11X across layers.
PolyScientist performance is close to the MKL-DNN's.
For some layers such as layer 11, PolyScientist is 9\% faster
while for a few layers notably layer 1, and 3, MKL-DNN performs better.
\emph{In this case of \textsf{fastrcnn}, we see that \textsf{PolyScientist}
outperforms \textsf{Microkernel} significantly clearly showing the need
for outer loop tuning in addition to having a high performance implementation 
of matrix multiplication in the inner most loops.}
\textsf{PolyScientist} picked code achieves \textasciitilde 2X performance gains over 
the code with the default loop order for layers 4, 7, 8, 10, and 11
while for layer 25, \textsf{PolyScientist} is 56\% higher performing.
Across all layers of \textsf{fastrcnn}, \textsf{PolyScientist} improves
the performance by 28\% on average.
Figure \ref{fig:fastrcnn_distro} shows the performance distribution for all
layers of \textsf{fastrcnn}. Here, we see that the performance distribution is great:
the difference between the performance of the best and the worst code variant 
seen is vast for all layers except layer 18. We observe that PolyScientist is
able to pick a variant whose performance is close to the performance of the 
best performing version.

From Figure \ref{fig:maskrcnn} through Figure \ref{fig:googlenetv3_distro}, we show the performance
achieved by various systems and the performance distribution of code variants
seen for all other CNN models, namely, \textsf{maskrcnn}, \textsf{xception},
\textsf{yolov2}, \textsf{mobilenet}, \textsf{alexnet}, \textsf{overfeat},
\textsf{googlenetv1}, and finally \textsf{googlenetv3}.
In Figure \ref{fig:maskrcnn} we observe that the performance of two layers
of \textsf{maskrcnn}
-- layer 31, and 32 is very low compared to the machine peak.
The reason is, the image sizes for the two layers are 7X7 and 1X1 respectively.
Consequently, the amount of work that each core has to perform is less and therefore,
MKL-DNN and PolyScientist are not able to attain performance close to the machine
peak.
For \textsf{maskrcnn} too, we discover that the default loop order -- \textsf{Microkerenel}, leaves a lot of performance on the table:
for layers 4, 5, 6, 9, 10, 14, 15, \textsf{PolyScientist} gets more than 2X extra 
performance compared to only the use of the microkernel. 
For layer 7, \textsf{PolyScientist} is 3X higher performing than \textsf{Microkernel}.
Across all layers of \textsf{maskrcnn}, the average performance benefit is
41\%.
In Figure \ref{fig:xception_distro}, we see that PolyScientist picks the right 
variant for all layers for \textsf{xception}.
For \textsf{yolov2}  from Figure \ref{fig:yolov2}, we note that
PolyScientist performance closely matches that of MKL-DNN and 
through Figure \ref{fig:yolov2_distro}, we see there is a great spread in 
performance of various code variants run.
In \textsf{mobilenet}, PolyScientist is as high performing as MKL-DNN 
(Figure \ref{fig:mobilenet}). Further, the different code variants 
perform very similarly for all layers of \textsf{mobilenet} (Figure \ref{fig:mobilenet_distro}).
In \textsf{alexnet}, we hardly see any difference in the relative performance
across the layers -- Figure \ref{fig:alexnet_distro}.
In \textsf{overfeat}, PolyScientist is slightly higher performing than MKL-DNN
and the performance spread is fair among different code variants generated
(Figures \ref{fig:alexnet}, and \ref{fig:alexnet_distro}).
\textsf{googlenetv1} and \textsf{googlenetv3} feature many more unique layers
and PolyScientist's performance is slightly better than MKL-DNN's for 
\textsf{googlenetv1} and is slightly worse for \textsf{googlenetv3} 
(Figures \ref{fig:googlenetv1}, and \ref{fig:googlenetv1}).

Table \ref{table:perf} shows the average performance of the four systems --
baseline, PolyScientist, PolyScientist-DNN, and MKL-DNN for various CNN models.
The performance of \textsf{PolyScientist-DNN} is consistently
slightly better than that of \textsf{PolyScientist}. 
Further, \textsf{PolyScientist} achieves magnitudes of higher performance
compared to the baseline code and 
is very competitive with respect to the hand crafted MKL-DNN library code.
In fact, PolyScientist outperforms MKL-DNN in the case of  
\textsf{alexnet}, \textsf{overfeat}, and \textsf{googlenetv1}.
\section{Related Work}
\label{sec:related}
Researchers have developed auto parallelization and program transformation systems to automatically transform code to achieve high performance \cite{uday08pldi,Kong:2019:MTM:3314221.3314653}. 
The state-of-the-art polyhedral compiler -- Pluto \cite{uday08pldi} derives a schedule
for the code that attempts to minimize data reuse distances.
The effect of the Pluto transformation will be that 
the iterations that use the same data will be executed close to each other in time
and therefore, it will be cache friendly.
However, Pluto's 
performance can be far from what we can achieve with the use of microkernels that exploit
the vector hardware effectively and by doing outer loop tuning
in the way we have developed this work.
Our initial experimental evaluation comparing our work with Pluto
revealed that its performance can be off by as much as 100X.
Kong et al. \cite{Kong:2019:MTM:3314221.3314653} develop a framework to decompose a program into sub-parts and use customized criteria (as opposed to using a single objective function) -- such as stride optimization, outer loop optimization, inner loop optimization to transform code. They show that their work without tiling transformation is able to achieve comparable results to that of Pluto.

Compile-time modeling of cache behavior and in particular calculating the number of cache misses
has been an active area of research \cite{ghosh1997cache,bao2017analytical,gysi2019fast}.
The researchers have demonstrated good accuracy in predicting the number of cache misses
on simulators. The modern computer architectures employ a hash based scheme to map
memory addresses to cache sets \cite{yarom2015mapping} which breaks
the assumptions behind the cache miss analysis. Consequently, their usefulness in modeling the performance of programs will suffer.
In the present work, we model the behavior of caches as well. 
However, we do not model cache misses rather we consider data reuses and determine
the size of cache needed to exploit the data reuses under conservative conditions.
We ignore streaming accesses as their misses in cache will not be crucial in the resulting
performance. Because of the these techniques, we show that we are able to accurately
rank code variants in terms of performance.
Latte \cite{Truong:2016:LLC:2908080.2908105} is a domain specific language and run time for 
coding of deep neural networks. Latte relies on pattern matching to transform
code patterns to library calls. For example, when a convolution operation is recognized,
a library call is made to the CPU-specific library, say Intel MKL-DNN.
Our work in this paper could be complementary to that of Latte where
instead of relying on an external hand coded library, Latte could be made to use PolyScientist
for high performance implementations of deep learning primitives. 

Strout et al \cite{Strout:1998:SSM:291069.291015} introduce
the novel concept of \emph{universal occupancy vector} (UOV)
for storage optimization --- expand the arrays to a minimal degree
so that the \emph{false} dependences are eliminated paving
the way for better scheduling of the code.
Thies et al \cite{Thies:2007:STU:1286821.1286825}
develop methods to find a good schedule when the storage mapping
is fixed and vice versa.
Ding et al  \cite{Ding:2003:PWL:780822.781159}
introduce approximate reuse distance analysis and sampling based
 profiling
to predict whether a given application exhibits regular data accesses
or irregular data accesses.
Xiang et al \cite{Xiang:2013:HHO:2499368.2451153}
develop a theory of locality and show how different locality metrics
are related to each other.

Autotuning systems using parametric tiling \cite{tavarageri2013adaptive,renganarayanan2007parameterized,darte2014parametric,baskaran2010parameterized,hartono2009parametric,tavarageri2010parametric}
perform tile size exploration with tiled code generated with tile sizes as parameters (rather than hard-coded tile sizes). The parametric tiling technology reduces the code generation
time during auto-tuning as the same code can be used for multiple tile size explorations.
AutoTVM \cite{chen2018learning} is targeted at accelerating deep learning workloads and uses machine learning to guide auto-tuning of deep learning primitives. 
Tiramisu \cite{baghdadi2019tiramisu} is a polyhedral model based compiler framework that introduces a scheduling language to allow the programmer to explore various
program transformations. 
Active Harmony \cite{chung2004using} determines the parameters that are critical for
performance among the set of parameters that need to be auto-tuned in a given application and focuses on optimizing them first in order to accelerate the process of
auto-tuning. CHiLL \cite{chen2007model} is a compiler framework that assists
an auto-tuner in generating code for schedules expressed in a high level language.
Tiwari et al \cite{tiwari2009scalable} combine ActiveHarmony and CHiLL to 
systematically explore the space of program transformations.
All of the above auto-tuning systems, while useful, incur a huge expense in terms of 
computation cost and time. The compiler algorithm developed in the paper 
narrows down the search space to only a small number of code variants thereby
saving the auto-tuning cost.

TVM \cite{chen2018tvm}, a compiler for deep learning, introduces the concept of \emph{tensorization}, where 
a unit of computation can be replaced with a microkernel written using hardware
intrinsics. The TVM tensorization tool can be complementary to our work -- 
our work can be leveraged within TVM to perform optimization of outer loops around
tensorized code, much like how we optimize code around
the microkernel in this paper. 

\section{Conclusion}
\label{sec:conclusion}
In this paper, we presented a novel cache data reuse algorithm.
We leverage the reuse analysis to perform \emph{relative ordering} of 
program variants. Using these techniques, we develop a framework 
to generate high performing code for deep learning primitives.
The framework also integrates the expert-coded microkernels,
further enabling the development of high performing code that achieves
 performance nearly identical to that of a hand-coded library.
The presented methodology will allow the creation of deep learning kernels
in an automated fashion easing the burden of writing hand-tuned libraries
and at the same time, realizing state-of-the-art efficiencies in utilizing modern
dense processing engines.

\balance
\bibliographystyle{ACM-Reference-Format}
\bibliography{paper}

\end{document}